\documentclass[final,floating,letterpaper]{siamltex1213}
\usepackage{amsmath,amssymb}
\allowdisplaybreaks[4]

\usepackage{comment}
\usepackage[compress]{cite}
\usepackage{algorithm}
\usepackage{algorithmic}

\algsetup{indent=2em}

\usepackage{graphicx}
\usepackage{epsfig,epsf,psfrag}
\usepackage{epstopdf}
\usepackage{bm}
\usepackage{hyperref}
\usepackage{amsmath,bm}
\usepackage{showkeys}

\title{Modified Poisson-Nernst-Planck model with Coulomb and hard-sphere correlations}

\author{Manman Ma
\thanks{School of Mathematical Sciences, Tongji University, Shanghai 200092, China ({\tt mamm@tongji.edu.cn}).}
\and
Zhenli Xu
\thanks{School of Mathematical Sciences, Institute of Natural Sciences,
  and MoE-LSC, Shanghai Jiao Tong University, Shanghai 200240, China ({\tt xuzl@sjtu.edu.cn}).}
\and
Liwei Zhang
\thanks{Institute of Natural Sciences, Shanghai Jiao Tong University, Shanghai 200240, China ({\tt zhangliwei01@sjtu.edu.cn}).}
}

\begin{document}

\maketitle

\begin{abstract}
We develop a modified Poisson-Nernst-Planck model which includes both the long-range Coulomb and short-range hard-sphere correlations
in its free energy functional such that the model can accurately describe the ion transport in complex environment and under a nanoscale
confinement. The Coulomb correlation {including the dielectric polarization} is treated by solving a generalized Debye-H\"uckel equation which is a Green's function equation
with the correlation energy of a test ion described by the self Green's function. The hard-sphere correlation is modeled through the
modified fundamental measure theory. The resulting model is available for problems beyond the mean-field theory such as problems with
variable dielectric media, multivalent ions, and strong surface charge density. We solve the generalized Debye-H\"uckel equation by the
Wentzel-Kramers-Brillouin approximation, and study the electrolytes between two parallel dielectric surfaces. {In comparison to
other modified models, the new model is shown more accurate in agreement with particle-based simulations and capturing the physical properties 
of ionic structures near interfaces.}
\end{abstract}



\pagestyle{myheadings}
\thispagestyle{plain}
\markboth{M. Ma, Z. Xu and L. Zhang}
{MPNP model with long- and short-range correlations}

\graphicspath{{Figs/}}

\section{Introduction}
Ion transport in nanoscale systems is of essential importance in many physical phenomena and biological processes,
and has attracting broad interest \cite{Schoch08RMP, French10RMP}. The Poisson-Nernst-Planck (PNP) model
is one of the most widely used models to describe the ion transport in fluids, where the dynamics of ions is composed
of the diffusion and the convection, and the convection is due to the gradient of the electric potential governed
by the Poisson's equation. Analytical and numerical analysis based on the PNP model has been successfully applied to
understand problems in areas such as nanofluidics, ion channels and semiconductor devices \cite{Bazant04PRE, Markowich90, Eisenberg12ACP, Eisenberg07SIMA}.

The classical PNP model is a mean-field theory as the electrostatic energy in the Nernst-Planck (NP) equations is the
mean potential energy. This model ignores ion-ion correlation, solvation energy due to variable dielectric
permittivity, dielectric boundary effects and the excluded volume effects, and thus for many instances it cannot
provide accurate description of particle interactions, significantly influencing the correct understanding of physical properties of
electrolytes. For example, when divalent ions are present in the electrolyte, the ions cannot be treated
as point charges without size effect, and with large surface charges the long-range Coulomb correlation cannot be ignored
in order to capture many-body phenomena such as charge inversion and counterion condensation effects.
Many efforts have been made on the improvement of the mean-field theory in literature \cite{Mamonov03BJ,Corry03BJ,KBA:PRE:2007, Eisenberg10JCP, Lu11BJ, Ji12JDDE, Frydel16}
in order to make the model more physical. {More recently,  the contributions from different components such as ion-ion correlations and excluded volume effects are systematically studied in Ref. \cite{voukadinova2019energetics}.}

To take into account the long-range Coulomb correlation, Bazant {\it et al.} \cite{Bazant11PRL} proposed a phenomenological
electrostatic free-energy functional treating ion-ion correlations by  Landau-Ginzburg-type theory, and this work has been
further developed in many studies \cite{Storey12PRE, Liu13JPCB, Giotti18SIAM,misra2019theory,de2020continuum}. Another promising approach to include the long-range
correlation is to correct the mean potential by the self energy of ions under the idea of the Gaussian variational field
theory \cite{podgornik1989jcp, NO:EPJE:2000, NO:EPJE:2003} where the self energy of a test ion can be defined by the diagonal of
a Green's function described by the generalized Debye-H\"uckel (GDH) equation and the many-body physics due to the ionic correlation
and dielectric image charges are both included \cite{XML:PRE:2014, Ma14JCP}. Beyond the point charge approximation, Wang \cite{Wang10PRE}
discussed the importance of the ionic self energy with finite size, developed a field-theoretical model with Gaussian charge
distribution to represent the test ion, which can be simplified to the Born-energy augmented Poisson-Boltzmann theory by
assuming the ionic radius being a small parameter for problems with variable dielectric permittivity.
Liu {\it et al.} \cite{Liu18SIAM} assumes the test ion being a hard sphere with a point charge at the center and obtains
the free energy functional by the Debye charging process using asymptotic expansion with the spherical radius as small parameter,
which includes the Coulomb correlation of ions with finite size and the dielectric effect, such that the total self energy is
described as the combination of the mean-field energy and correlation self energy where the Born solvation energy is the leading
contribution for variable media. This type of modified PNP models satisfies the energy dissipation law and a proper weak formulation,
and has been successfully applied in exploring important physical properties, such as like-charge attraction
and double layer structure \cite{Liu17CCP, Ji18SIAP}. {As an alternative approach,  the density functional theory has also been extensively studied in the past few decades where long-range Coulomb correlation by integral equation theory such as 
hypernetted-chain and mean spherical approximations \cite{kierlik1991density,rosenfeld1993free,gillespie2003density,roth2016shells}}.

Another key factor for the ion transport is the steric effect of ions. Though the finite size is included in the model
of Coulomb correlation, the short-range non-electrostatic correlation is not accounted for.
For simple systems of hard-sphere (HS) fluids, short-range interactions are generally repulsive arising from the excluded volume effect
of particles, more precisely, the Pauli exclusion that prevents two electrons from occupying the same quantum state.
Among various models for implementing the HS interactions, the density functional theory has been widely utilized as
a powerful approach by capturing the non-local contributions to the free energy functional, where the fundamental measure theory (FMT)
proposed by Rosenfeld \cite{Rosenfeld89PRL} has been shown to investigate the HS systems successfully. There are many subsequent work
to make the improvements of the FMT from different aspects. It is widely accepted now that the steric interaction for the HS model
can be accurately calculated by the modified fundamental measure theory (MFMT) \cite{Roth:JPC:2002,Yu2002JCP} for the excess free energy functional.
In the FMT, the excess Helmholtz energy functional can be expressed as a weighted density approximation which can
provide a more accurate description of properties of inhomogeneous HS fluids and becomes the same as that
from the scaled particle theory \cite{Reiss59JCP} or the Percus-Yevick theory \cite{Lebowitz65JCP}
in the limit of homogeneous hard spheres. The original FMT overestimates the contact value of density distribution and shows evident deviations under strong confinements in comparison to Monte Carlo simulations.
Yu and Wu \cite{Yu2002JCP} {and Roth {\it et al.} \cite{Roth:JPC:2002}} proposed the MFMT by using a more accurate Helmholtz energy density
from the Boublik-Mansoori-Carnahan-Starling-Leland equation of state
for the HS correlation. Further, Jiang {\it et al.} \cite{Jiang14JPCM} generalized it to transport problems and obtained the time-dependent density functional theory
using the framework of the NP equations. Separately, there are also many other theories describing the excess free energy of steric effects such as
those based on the Lennard-Jones kernel\cite{Lin05CPAM, Sheng08PTPS, Eisenberg10JCP}, entropic effect of solvent molecules \cite{Borukhov97PRL, Li09nonlinearity, Li09SIMA, Li13Nonlinearity, Lu11BJ},
and Carnahan Starling local density\cite{Giera13PRE}.

{Besides contributions to the continuum theory, there have been plenty of results based on particle-based simulations 
for primitive electrolytes and they are in general thought as benchmarks to different theoretical models due to proper 
treatments of Coulombic correlations and steric effects. For instance, the planar system without dielectric mismatches 
is well discussed by a recent 
systematic Monte Carlo (MC) study \cite{valisko2018systematic} and references therein.
For problems with dielectric interfaces that is related to current work, we refer to MC studies  \cite{boda2004computing,henderson2005monte,nagy2011simulation,dos2015electrolytes,wang2012molecular}
 and recent molecular dynamic (MD) simulations \cite{liang2020harmonic}. }

In this work, we propose a new modified PNP model for the ion transport, which contains
both the long-range Coulomb and the short-range HS correlations. In the model, the total energy of an ion in the system
consists of three contributions, the mean-field electrostatic energy by the Poisson's equation, the Coulomb self energy by the
GDH equation and the excess Helmholtz energy from the MFMT. 
The Wentzel-Kramers-Brillouin (WKB) approach is proposed for
problems with two parallel dielectric interfaces and the contribution of the Stern layer is accounted for in this
approach, which is essential to the accuracy of the approximation. By the WKB approximation, the modified PNP model
can be efficiently solved by finite-difference numerical methods. 
{Comparisons among classical mean-field PNP, and
different versions of modified PNP equations are made, together with particle-based simulations}. 
Our numerical simulations demonstrate that the new model can accurately
predict the oscillation of the ionic density as well as depletion zone near highly charged dielectric interfaces. 
Furthermore, the combined effects of dielectric boundary and ionic size  in current model are well presented and discussed through simulations with various parameter groups.

The rest of the paper is organized as follows. In Sec. \ref{MPNP}, we formulate the modified PNP model with including long-
and short-range correlations through an energetic approach. In Sec. \ref{1Dmodel}, a two-plate model together with its dimensionless form
is described and the method for solving the modified PNP system is proposed. In Sec. \ref{num}, numerical calculations are
performed to show the attractive feature of the model. The conclusion remarks are finally made in Sec. \ref{con}.

\section{Modified Poisson-Nernst-Planck model}\label{MPNP}

We consider the ionic dynamics in electrolytes composed of $K$ ionic species. Let $c_i(\bm{r};t)$ be the ionic density of the $i$-th species
at time $t$ and $z_i$ be its valence. The evolution of the ionic densities obeys the general conservation law,
\begin{equation}\label{conservation}
	\frac{\partial c_i}{\partial t} = -\nabla\cdot \bm{J}_i,
\end{equation}
where the ionic flux can be described by the constitutive relation with the electrochemical potential $\bar{\mu}_i$,
\begin{equation}\label{flux}
	\bm{J}_i=-b_ic_i\nabla\bar{\mu}_i,
\end{equation}
with $b_i$ being the ionic mobility of the $i$-th species. With the ionic density given, the distribution of
electric potential $\phi$ over the space  is governed by the Poisson's equation,
\begin{equation}
	-\nabla\cdot\varepsilon\nabla \phi = \rho_f+\sum_{i=1}^K z_iec_i,\label{eq:Poisson}
\end{equation}
where $\varepsilon$ is the dielectric permittivity of the medium, $\rho_f$ is the fixed charge density in space,
and $e$ is the elementary charge. The dielectric permittivity is a physical quantity that could depend on position,
electric field or ionic concentrations \cite{B:JCP:1951,B:JCP:1955,PCZ:JCP:87,TCS:PRE:01,BGN:PRL:2011,BAP:JCP:11,LWZ:CMS:14},
and this work assumes a space-dependent function $\varepsilon(\bm{r})$.

To connect the electrochemical potential with the electric potential and ionic densities, we consider an energetic approach
by a variation of the total free energy $F$ with respect to the ionic density, that leads to
$\bar{\mu}_i=\delta F/\delta c_i$.
We assume that the charged system in domain $\Omega$ is composed of ions represented by charged hard-spheres.
In this work, the total free energy is decomposed into three contributions, expressed by the following formula,
\begin{equation}\label{totalenergy}
	F = F^{\textup{mf}} + F^{\textup{co}} + F^{\textup{hs}},
\end{equation}
where $F^{\textup{mf}}$ is the electrostatic free energy of mean-field approximation, $ F^{\textup{co}}$ is 
the Coulomb correlation energy, and $F^{\textup{hs}}$ describes the excess free energy due to HS repulsive interactions.

The free energy is a functional of the ionic density vector $\bm{c}=(c_1,\cdots, c_K)$.
The mean-field electrostatic free energy has been widely studied and is given by,
\begin{align}\label{Fmf}
\begin{split}
	 F^\textup{mf}[\bm{c}] = \int_\Omega \left[\frac{\varepsilon(\bm{r})}{2}|\nabla\phi(\bm{r})|^2 + k_BT\sum_{i=1}^K c_i\left(\ln \Lambda^3 c_i -1\right) \right]d\bm{r},
\end{split}
\end{align}
where the potential is the solution of the Poisson's equation together with some boundary conditions,
$k_B$ is the Boltzmann constant, $T$ is the absolute temperature {and $\Lambda$ is the thermal de Broglie wavelength. The chemical potential due to this functional is,
\begin{equation}
	\mu_i^\textup{mf} = \frac{\delta F^\textup{mf}}{\delta c_i} = z_ie\phi + k_BT\ln c_i.
\end{equation}
When $\bar{\mu}_i$ is replaced by this chemical potential, one obtains the classical NP equations from Eq. \eqref{conservation},
and thus the PNP equations coupled with the Poisson's equation \eqref{eq:Poisson}.
At equilibrium, the density  minimizes the free energy, that gives straightforwardly a constant chemical potential.
This results in the classical Boltzmann distribution to represent the ionic density as function of $\phi$, and leads
to the mean-field Poisson-Boltzmann equation,
\begin{equation}
	-\nabla\cdot\varepsilon\nabla \phi = \rho_f+\sum_{i=1}^K z_ie c_i^{\textup{b}}\exp{\left(-\beta z_ie \phi\right)},
\end{equation}
where $\beta=1/(k_BT)$ and $c_i^{\textup{b}}$ is the bulk-state ion density.

As discussed in the Introduction, the PNP and Poisson-Boltzmann models are mean-field theories and cannot
capture many-body effects, and to go beyond correlations should be taken into account.
The excess free energies due to long-range Coulomb and short-range HS correlations, $F^{\textup{co}}$ and
$F^\textup{hs}$ as functionals of the ionic densities will be discussed separately below, which
result in the corresponding components for chemical potentials, $\mu_i^\textup{co}=\delta F^{\textup{co}}/\delta c_i$
and $\mu_i^\textup{hs}=\delta F^{\textup{hs}}/\delta c_i$. The total electrochemical potential is
$	\bar{\mu}_i =\mu_i^\textup{mf} + \mu_i^\textup{co} + \mu_i^{\textup{hs}}$.
Finally, by Eqs. \eqref{conservation} and \eqref{flux}, one obtains the modified NP equations,
\begin{equation}\label{mNP}
	\frac{\partial c_i}{\partial t} = \nabla\cdot \left[ D_i\nabla c_i +b_i c_i\nabla \left(z_ie\phi+\mu_i^\textup{co}+\mu_i^\textup{hs}\right) \right] ,\quad i=1,...,K,
\end{equation}
where $D_i$ is the ion diffusivity connecting the ionic mobility by the Einstein relation $D_i = k_BTb_i$.
Alternatively, one can denote by
	$U_i = z_ie\phi+\mu_i^\textup{co}+\mu_i^\textup{hs}$
the potential of mean force that means the energy cost to move a single ion from bulk to the current position.
Together with the Poisson's equation \eqref{eq:Poisson}, the coupled system is named as the modified
PNP equations and is self-consistent because $\mu_i^\textup{co}$ and $\mu_i^\textup{hs}$ are functions
depending on ion densities only,  as shown in Sec. \ref{sec:coulomb} and \ref{sec:hs}.

\subsection{Coulomb correlation} \label{sec:coulomb}


{The long-range Coulomb correlation is the long-range electrostatic correlation, including both the ion-ion correlation beyond mean field} and the dielectric self energy of ions \cite{Liu18SIAM}.
By the Debye charging process, the Coulomb correlation can be written as,
\begin{equation} \label{eq:cc}
	F^\textup{co}  = \int d\bm{r} \int_0^1 d\lambda^2\sum_{i=1}^K c_i(\bm{r})\left[U^{\textup{Born}}_i(\bm{r})+U_i^{\lambda}(\bm{r})\right],
\end{equation}
where $\lambda\in[0,1]$ is increased from 0 to 1, that represents the charging process, and
$U^{\textup{Born}}_i(\bm{r})$ is the Born solvation energy \cite{Born:ZP:1920} which is space-dependent for an inhomogeneous dielectric medium.
The second term in the bracket of Eq. \eqref{eq:cc} is the $\lambda$-dependent part due to  the
Coulomb correlation, which by the point charge approximation is given by,
\begin{equation} \label{eq:Ulam}
	U_i^{\lambda}(\bm{r}) = \frac{1}{2}\sum_{j=1}^K\int_\Omega c_j(\bm{r}')h_{ij}^{\lambda}(\bm{r},\bm{r}')v_{ij}(\bm{r},\bm{r}')d\bm{r}',
\end{equation}
where
\begin{equation} \label{eq:v_el}
	v_{ij}(\bm{r},\bm{r}') = z_iz_je^2G_0(\bm{r},\bm{r}'),
\end{equation}
is the pairwise potential contribution arising from the electrostatic interaction,
with the Green's function $G_0$ satisfying
\begin{equation} \label{eq:G0}
	-\nabla\cdot\varepsilon(\bm{r})\nabla G_0(\bm{r},\bm{r}') = \delta(\bm{r}-\bm{r}'),
\end{equation}
and $h_{ij}^\lambda(\bm{r},\bm{r}')$ is the $\lambda$-dependent pair correlation function.

The remaining step is to obtain the pair correlation function $h^\lambda_{ij}$ that
can be determined by the Ornstein-Zernike equation \cite{HM::2006},
\begin{equation} \label{eq:OZ}
	h^\lambda_{ij}(\bm{r},\bm{r}') = c^\lambda_{ij}(\bm{r},\bm{r}') + \int\sum_{k=1}^K c_k(\bm{r}'')h^\lambda_{jk}(\bm{r}',\bm{r}'')c^\lambda_{ik}(\bm{r},\bm{r}'')d\bm{r}''.
\end{equation}
In order to close the system, one applies the random phase approximation for the direct correlation function $c^\lambda_{ij}$, i.e.,
$c^\lambda_{ij} = -\beta\lambda^2 v_{ij}$ \cite{FM:PRE:2016}.
Similarly to the expression of Eq. \eqref{eq:v_el}, one can write
\begin{equation}
	h^\lambda_{ij}(\bm{r},\bm{r}')  = -\beta z_iz_je^2h_\lambda(\bm{r},\bm{r}').
\end{equation}
Together with Eq.  \eqref{eq:v_el}, one substitutes the expressions of $h^\lambda_{ij}$ and $c^\lambda_{ij}$
into the Ornstein-Zernike equation, and obtains,
\begin{equation} \label{eq:OZRPA}
	-h_\lambda(\bm{r},\bm{r}') = \lambda^2 \int G_0(\bm{r},\bm{r}'')\left[2I(\bm{r}'')h_\lambda(\bm{r}',\bm{r}'')-\delta\left(\bm{r}'-\bm{r}''\right)\right]d\bm{r}'',
\end{equation}
with the ionic strength expressed by
\begin{equation*}
	I(\bm{r}) = \frac{1}{2}\beta e^2\sum_{i=1}^K z_i^2c_i(\bm{r}).
\end{equation*}
Thus, one can write the $\lambda$-dependent part of the correlation energy  as
\begin{align}
	U_i^\lambda(\bm{r}) &= -z_i^2e^2\int G_0(\bm{r},\bm{r}')I(\bm{r}')h_\lambda(\bm{r},\bm{r}')d\bm{r}'\label{selfenergy}\\
	&=\frac{1}{2}z_i^2e^2\lim_{\bm{r}'\rightarrow\bm{r}} \left[\frac{h_\lambda(\bm{r},\bm{r}')}{\lambda^2} - G_0(\bm{r},\bm{r}')\right] \nonumber\\
	&= \frac{1}{2}z_i^2e^2\lim_{\bm{r}'\rightarrow\bm{r}} \left[G_\lambda(\bm{r},\bm{r}') - G_0(\bm{r},\bm{r}')\right] \nonumber
\end{align}
where one has denoted by the $\lambda$-dependent Green's function $G_\lambda = h_\lambda/\lambda^2$ that can be determined
by the differential equation transformed from Eq. \eqref{eq:OZRPA}, i.e., the generalized Debye-H\"{u}ckel equation,
\begin{equation} \label{eq:Glam}
	-\nabla\cdot\varepsilon(\bm{r})\nabla G_\lambda(\bm{r},\bm{r}') + 2\lambda^2I(\bm{r}) G_\lambda (\bm{r},\bm{r}') = \delta(\bm{r}-\bm{r}').
\end{equation}

The excess chemical potential coming from the Coulomb correlation is now derived as
\begin{equation}	\label{eq:mu_el}
	\mu_i^\textup{co}(\bm{r}) =\frac{\delta F^\textup{co}}{\delta c_i} = U_i^\textup{Born}(\bm{r}) + \frac{\delta}{\delta c_i} \int d
	\bm{r} \int_0^1 d\lambda^2\sum_{j=1}^K c_j(\bm{r})U_j^{\lambda}(\bm{r}).
\end{equation}
The integration with respect to $\lambda$ can be performed analytically. For the purpose,
we use Eqs. \eqref{eq:OZRPA} and \eqref{selfenergy} to expand $U_j^\lambda(\bm{r})$ as,
\begin{equation} \label{eq:uj}
	U_j^\lambda(\bm{r}) = \frac{z_j^2e^2}{2}\sum_{k=1}^\infty (-1)^{k}\lambda^{2k}\beta_k(\bm{r}),
\end{equation}
with $\beta_k$ the integral of form
\begin{equation*}
	\beta_k(\bm{r}) = \int d\bm{r}_1\cdots\int d\bm{r}_k\left[G_0(\bm{r},\bm{r}_1)2I(\bm{r}_1)\cdots G_0(\bm{r}_{k-1},\bm{r}_k)2I(\bm{r}_{k})G_0(\bm{r}_k,\bm{r})\right].
\end{equation*}
Thus, {with Eq. \eqref{eq:uj} plugged in,} the last term in Eq. \eqref{eq:mu_el} can be reformulated into
\begin{align}
	& \frac{1}{\beta}\int_0^1 d\lambda \frac{\delta}{\delta c_i}\int d\bm{r} \left[2I(\bm{r})\sum_{k=1}^\infty (-1)^{k}\lambda^{2k+1}\beta_k(\bm{r}) \right] \nonumber\\
	 = &\frac{z_i^2e^2}{2}\int_0^1 d\lambda \sum_{k=1}^\infty 2(k+1)(-1)^{k}\lambda^{2k+1}\beta_k(\bm{r})  \nonumber\\
	 = & \int_0^1 d\lambda \frac{d \left(\lambda^2  U_i^\lambda(\bm{r})\right)}{d\lambda} \nonumber\\
	 = &U_i^{\lambda=1}(\bm{r}),
\end{align}
where we have utilized the property,
$$\frac{\delta }{\delta c_i} \left(\int  2I(\bm{r})\beta_k(\bm{r})d\bm{r}\right)= \beta z_i^2e^2(k+1)\beta_k(\bm{r}).$$
Clearly, we have that the free energy contribution is exactly the correlation energy at the full charging.

For the ease of asymptotic analysis, the chemical potential is further rearranged as
\begin{align}
	\mu_i^\textup{co} (\bm{r})
	=& U_i^\textup{Born}(\bm{r}) + U_i^{\lambda=1} (\bm{r})\nonumber\\
	=& 	\left[ U_i^\textup{Born}(\bm{r})+\frac{1}{2}z_i^2e^2\lim_{\bm{r}'\rightarrow\bm{r}}\left(\frac{1}{4\pi\varepsilon(\bm{r}) |\bm{r}-\bm{r}'|}- G_0(\bm{r},\bm{r}')\right) \right]\nonumber \\
	& \quad \quad + \frac{1}{2}z_i^2e^2\lim_{\bm{r}'\rightarrow\bm{r}}\left[G_{\lambda=1}(\bm{r},\bm{r}') -  \frac{1}{4\pi\varepsilon(\bm{r}) |\bm{r}-\bm{r}'|} \right]. \label{eq:cochem}
\end{align}
We now apply the strategies of asymptotic expansions from Ref. \cite{Liu18SIAM} (Sec. 2.3.2) to the part within the first bracket in Eq. \eqref{eq:cochem} and have its leading order term, i.e., the $O(1/a_i)$ term, as the following Born solvation energy,
\begin{equation}	\label{eq:born}
	 \frac{z_i^2e^2}{8\pi a_i}\left(\frac{1}{\varepsilon(\bm{r})} - \frac{1}{\varepsilon_0} \right),	
\end{equation}
with $a_i$ the radius of the $i$th ion. It should be noted that using ion radius in \eqref{eq:born} is an approximation and could lead to a poor prediction sometimes as compared to experiments because the ion and Born radii differ and produce very different Born energies for real electrolytes.
The point charge assumption involved in Eq. \eqref{eq:Ulam} has generated the decoupling of the size effect embodied in Born energy \eqref{eq:born} and the ion-ion interactions related to $G_\lambda$ as described by Eq. \eqref{eq:Glam}. For more general cases of size-dependent $G_\lambda$ and higher-order correction to the Born energy, we refer
systematical analysis to the asymptotic work in Ref. \cite{Liu18SIAM}.

It is noted that there is another way used to obtain the chemical potential, called the G\"untelberg charging process \cite{guntelberg1926}
where only the central charge of the test ion is increased from zero to the full charging and other ions remain full charges during
the process. It was shown \cite{Liu18SIAM} that the Debye and G\"untelberg charging processes are equivalent in the case of
the point charge approximation. It is then straightforward to have the following expression for the Coulomb correlation component of the chemical potential, 
\begin{equation}\label{electromu}
	\mu^{\textup{co}}_i=\frac{z_i^2e^2}{2}\left[\frac{1}{4\pi a_i}\left(\frac{1}{\varepsilon(\bm{r})} - \frac{1}{\varepsilon_0} \right) +\lim_{\bm{r}'\rightarrow \bm{r}} \left(G(\bm{r},\bm{r}')-\frac{1}{4\pi\varepsilon(\bm{r}) |\bm{r}-\bm{r}'|}\right)\right],
\end{equation}
where $G(\bm{r},\bm{r}')=G_{\lambda=1}(\bm{r},\bm{r}')$ is the Green's function satisfying the GDH equation
with full charging,
\begin{equation}\label{eq:GDH2}
	-\nabla\cdot\varepsilon(\bm{r})\nabla G(\bm{r},\bm{r}') + 2I(\bm{r})G(\bm{r},\bm{r}') = \delta(\bm{r}-\bm{r}').
\end{equation}
It is mentioned that from now on we have omitted $t$ in notations for time-dependent  ionic densities and related variables, for instance, $c_i(\bm{r}) = c_i(\bm{r}; t)$ and $G(\bm{r},\bm{r}') = G(\bm{r},\bm{r}'; t)$.
In order to involve the boundary effects, one needs to define the ionic strength profile in general as
\begin{eqnarray}
    I(\bm{r}) &&= \left\{ \begin{array}{ll}
                \frac{1}{2}\beta e^2\sum_{i=1}^K z_i^2c_i(\bm{r}),~~\bm{r}\in\Omega, \\
                 0,~~~~~~~~~~~~~~~~~~~~~~~~\bm{r}\notin\Omega,
\end{array}\right.
\end{eqnarray}
and consider the  inhomogeneous dielectric permittivity profile.
Noteworthily, the GDH equation \eqref{eq:GDH2} has been applied to the exterior of the electrolyte besides taking into account the ion-ion interactions within the electrolyte, that means the dielectric property and absence of ions in the boundary domain, or equivalently, the dielectric boundary effects are effectively implemented in our model.

\subsection{Hard-sphere correlation} \label{sec:hs}
In addition to long-range Coulomb correlations, there are also non-electrostatic HS interactions that are generally repulsive
and short-ranged. Various models have been developed for the HS interactions such as FMT \cite{Rosenfeld89PRL}
and its modifications \cite{Roth:JPC:2002,Yu2002JCP}. This work uses the MFMT which accurately approximates the excess Helmholtz energy density
and has been shown promising to provide an efficient implementation of HS interactions \cite{Roth:JPC:2002,Yu2002JCP,Roth2010JPCM,Zhao2015ACE}.
Since the MFMT has been discussed in many literatures, here we present the basic idea of the theory.

In the MFMT, the excess Helmholtz free energy due to HS interactions can be expressed by a weighted density approximation,
\begin{equation} \label{eq:Fhs}
	F^\textup{hs}[\bm{c}] = k_BT\int f^{\textup{hs}}[\bm{c}(\bm{r})]d\bm{r},
\end{equation}
where  the excess Helmholtz energy density  $f^\textup{hs}$ is written as a function of six weighted densities which are
functionals of the ionic densities, including four scalar densities $n_\alpha(\bm{r}), (\alpha=0,\cdots,3),$ and
two vector densities $\bm{n}_{a}(\bm{r})$ and $\bm{n}_{b}(\bm{r})$, precisely,
\begin{align}
	f^\textup{hs}[\bm{c}(\bm{r})] =& -n_0\ln(1-n_3) + \frac{n_1n_2-\bm{n}_a\bm{n}_b}{1-n_3} \nonumber\\
	&+ \frac{(1-n_3)^2\ln(1-n_3)+n_3}{36\pi n_3^2(1-n_3)^2}\left(n_2^3-3n_2\bm{n}_b\cdot\bm{n}_b\right). \label{eq:fhs}
\end{align}
The weighted densities are defined by the weighted averages of ionic densities,
\begin{equation}
	n_\alpha (\bm{r}) =  \sum_{i=1}^{K}n^{(\alpha)}_i (\bm{r}) =  \sum_{i=1}^{K} \int c_i(\bm{r}')\omega_{i}^{(\alpha)}(\bm{r}-\bm{r}')d\bm{r}',\label{eq:na}
\end{equation}
where the relevant six weight functions are classified into four scalar weights,
\begin{align}
\omega_{i}^{(0)}(\bm{r}) & =\frac{\omega_{i}^{(1)}(\bm{r}) }{a_i}= \frac{\omega_{i}^{(2)}(\bm{r})}{4\pi a_i^2} = \frac{\delta(a_i-r)}{4\pi a_i^2},\quad
\omega_{i}^{(3)}(\bm{r})= \theta(a_i-r),
\end{align}
and the two vector-type weights,
\begin{equation}
	\bm{\omega}_{i}^{a}(\bm{r})  =\frac{\bm{\omega}_{i}^{b}(\bm{r})}{4\pi a_i} = \frac{ \bm{r}}{r}\frac{ \delta(a_i-r)}{4\pi a_i}.
\end{equation}
Here $r=|\bm{r}|$, 
$\delta(\cdot)$ and $\theta(\cdot)$ are the
Dirac delta and the Heaviside step functions, respectively. It can be seen that $\omega_{i}^{(1)}$ and $\omega_{i}^{(2)}$ are
proportional to $\omega_{i}^{(0)}$, $\bm{\omega}_{i}^{b}$ is proportional to $\bm{\omega}_{i}^{a}$, and thus
there are actually only two scalar weights and one vector weight. Eventually, $F^\textup{hs}$ turns out to be a functional of ion densities only.
As a consequence, the variational derivative of excess free energy with respect to the ionic density
leads to the corresponding HS excess chemical potential,
where the derivatives of energy density  to weighted densities can be  obtained from Eq. \eqref{eq:fhs} straightforwardly.

\section{Two-plate problem} \label{1Dmodel}
In this section, one considers the solution of a two-plate problem with an electrolyte placed in between two parallel
electrodes by solving the modified PNP model. This problem is often used in the understanding of the ionic structure near interfaces,
and is of importance in many electrochemical devices such as charge storages and water
purification.
 \begin{figure}[h!]
 \centering
 \includegraphics[width=0.7\linewidth]{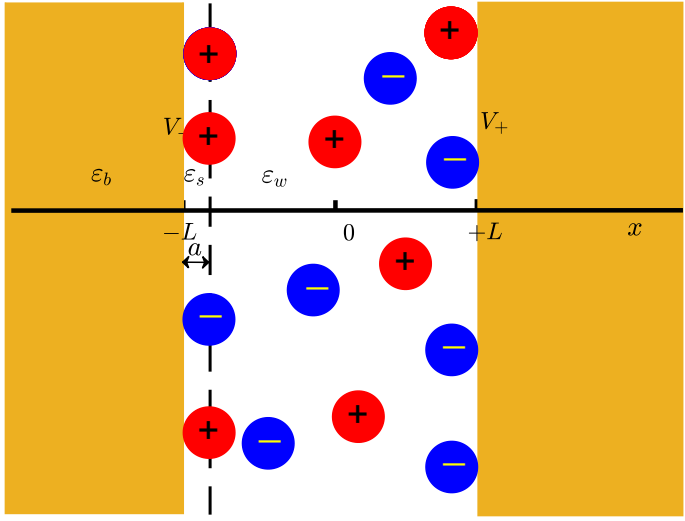}
 \caption{Schematic representation of a two-plate problem. Ions are inaccessible to the
 Stern layers of thickness $a$. Separation of two interfaces is $2L$ and dielectric coefficient
 is piecewise-constant with $\varepsilon_s$ in the Stern layers, different from the ion-accessible
 region and the electrode region.}
 \label{fig:sch}
 \end{figure}

\subsection{Dimensionless equations}
We assume that two interfaces are located at $x=\pm L$ with voltages $V_\pm$ (as shown in Fig. \ref{fig:sch}).
The electrolyte region near each electrode surface is divided into
an ion-impenetrable Stern layer with thickness  $a=\min_i{a_i}$ and a diffuse layer of mobile ions.  {Note that the dielectric profile within the electrolyte could be generally inhomogeneous, for instance, space-dependent near surfaces, concentration- or electric field-dependent. However, to simplify the model setup,}
the inhomogeneous dielectric profile of the entire domain is assumed to be homogeneous along
$yz$ directions, and is described by the piecewise function
$\varepsilon(x)$ which is $\varepsilon_s$ in the Stern layers $L-a\le |x|\le L$, $\varepsilon_w$
in the ion-accessible region $|x|<L-a$, and $\varepsilon_b$ in the electrode region $|x|> L$.
The modified PNP equations for the ion-accessible region are given by,
\begin{align}\label{1Dmpnp}
	&-\varepsilon_w  \frac{\partial^2 \phi}{\partial x^2} = \sum_i   z_iec_i, \\
	&\frac{\partial c_i}{\partial t} =\frac{\partial}{\partial x}\left[ D_i\frac{\partial}{\partial x}c_i+
b_ic_i\frac{\partial}{\partial x}\left(z_ie\phi+\mu^{\textup{co}}_i + \mu^{\textup{hs}}_i \right)\right].
\end{align}
The weighted densities associated with the HS correlation are simplified as one dimensional integrals,
\begin{align}\label{ni}
	&n_{i}^{(0)}(x)=\frac{1}{2 a_i}\int_{x-a_i}^{x+a_i}c_i(x')dx', \\
	&n_{i}^{(3)}(x)= \pi\int_{x-a_i}^{x+a_i}c_i(x')[a_i^2-(x'-x)^2]dx',\\
	&{\bm{n}}_{i}^{a}(x) = \frac{\bm{e}_x}{2a_i}\int_{x-a_i}^{x+a_i}c_i(x')(x'-x)dx',
\end{align}
and $n_{i}^{(1)},  n_{i}^{(2)}$ and $\bm{n}_i^b$ are defined consequently.

Simultaneously, the GDH equation associated with $\mu^{\textup{co}}_i$ is reduced to
\begin{equation}
	-\nabla\cdot\varepsilon(x)\nabla G  + 2 I(x)G = \delta(\bm{r}-\bm{r}'). \label{eq:Green}
\end{equation}
Note that the Green's function is three-dimensional with the geometrical symmetry although the dielectric and ionic strength profiles
are simply one-dimensional, i.e., $G=G(x,x',\rho)$ with $\rho = \sqrt{(y-y')^2+(z-z')^2}$. Below we focus on the asymptotic treatment of
its approximation for the solution of the GDH equation.

At interfaces between the Stern and diffuse layers $ x=\pm(L-a)$, proper boundary conditions have to be applied to solve modified PNP.
Precisely, the Laplace equation for $\phi$ within the Stern layer gives the Robin boundary conditions,
\begin{equation}
	\phi\pm a\frac{\varepsilon_w}{\varepsilon_s}\frac{\partial\phi}{\partial x} = V_\pm.
\end{equation}
Additionally, we consider the no surface-reactions assumption that leads to the no-flux boundary conditions,
\begin{equation}
	\bm{J}_i\cdot \bm{n} = D_i\left(\frac{\partial c_i}{\partial x}+\beta c_i\frac{\partial U_i}{\partial x}\right) = 0.
\end{equation}
Finally, the far-field boundary condition for Green's function reads
$	\lim_{\bm{r}\rightarrow\infty}G(\bm{r},\bm{r}') = 0.$
These boundary conditions together with the initial data of the ionic concentrations uniquely determine the solution
of the modified PNP system.

As a further step, the quantities in the system are nondimensionalized by using
$\varepsilon_w$ for dielectric permittivity, the bulk density $c_0$ for ion densities, $1/(\beta e)$ for potentials, $L$ for length, the uniform $D_0(=D_i)$ for diffusivities, $L\ell_0/D_0$ for time with the {length scale defined by $\ell_0=1/\sqrt{8\pi\ell_\textup{B} c_0}$ which is the Debye length for a 1:1 electrolyte} and the Bjerrum length $\ell_\textup{B}=\beta e^2/(4\pi\varepsilon_w)$.

For electrostatic-related variables, $G$  and $\mu_i^{\textup{co}}$  are rescaled by $1/(\beta e^2)$ and $1/\beta$, respectively, that gives the dimensionless chemical potential,
\begin{equation}
	\mu_i^{\textup{co}}=\frac{1}{2}z_i^2\left[ \frac{q}{a_i}\left(\frac{1}{\eta(x)} - \frac{1}{\eta_0}\right)+\lim_{\bm{r}'\rightarrow \bm{r}} \left(G(\bm{r},\bm{r}')-\frac{q	}{\eta(x)|\bm{r}-\bm{r}'|}\right)\right],
\label{eq:co}\end{equation}
where {$\eta$ is the rescaled dielectric profile}, the dimensionless parameter is $q=\ell_\textup{B}/L$,
and $G$ satisfies the dimensionless GDH equation,
\begin{equation} \label{eq:GDHdimless}
	-\nabla\cdot\eta(x)\nabla G + \frac{I(x)}{\epsilon^2}G = 4\pi q\delta(\bm{r}-\bm{r}').
\end{equation}
Here comes the second dimensionless parameter $\epsilon=\ell_0/L$,
and the spatial dependent dielectric and ionic strength profiles are
\begin{equation}
\eta(x) = \left\{ \begin{array}{ll}
                1, \quad|x|<1-a,\\
                \eta_s, \quad 1-a\le |x|\le 1, \\
                \eta_b, \quad |x|> 1.
\end{array}\right.
\quad
	   I(x) = \left\{ \begin{array}{ll}
                \sum_i z_i^2c_i(x)/2, \quad |x|<1-a,\\
               0, \quad \quad \quad \quad\quad\quad |x|\ge 1-a.
\end{array}\right.
\end{equation}

At the HS side, $\mu_i^{\textup{hs}}$ is rescaled by $1/\beta$ and the weighted densities $n_\alpha$ are rescaled by
$\nu = c_0L^3 = 1/(8\pi q\epsilon^2)$. Consequently, the dimensionless integrals transformed from the variation of Eq. \eqref{eq:Fhs} can be evaluated directly though detailed formulae are omitted here.

Finally, we obtain the dimensionless form of the modified PNP,
\begin{align}\label{dimensionlessmpnp}
&-2\epsilon^2\frac{\partial^2}{\partial x^2}\phi = \sum_{i=1}^K z_ic_i,\quad |x|\le 1-a, \\
	&\frac{\partial c_i}{\partial t} = \epsilon\frac{\partial}{\partial x}\left[ \frac{\partial}{\partial x}c_i+c_i\frac{\partial}{\partial x}\left(z_i\phi+\mu^{\textup{co}}_i + \mu^{\textup{hs}}_i \right)\right],\quad |x|\le 1-a,
\end{align}
with boundary conditions
\begin{align}
	&\phi\pm a\frac{1}{\eta_s}\frac{\partial\phi}{\partial x} = V_\pm,\quad x=\pm(1-a), \\
	& \frac{\partial c_i}{\partial x} +c_i\frac{\partial }{\partial x}(z_i\phi+\mu^{\textup{co}}_i + \mu^{\textup{hs}}_i)= 0, \quad x=\pm(1-a).
\end{align}

As an important quantity in investigating electric double layers,  the total diffuse charge near one electrode is also considered. The dimensionless value of that, without loss of generality, near the negative-charged electrode is
\begin{equation}
	Q(t) = \int_{-1+a}^0 \sum_{i=1}^Kz_ic_i(x,t) dx.
\end{equation}

\subsection{WKB approximation}
The Green's function involved in the GDH equation is generally six-dimensional and dependent on both $\bm{r}$ and $\bm{r}'$.
Due to the symmetry in the geometry, the system can be simplified by Fourier transform and reduced to a two-dimensional equation
that can be solved efficiently by selected inversion method \cite{XuMaggs:JCP:14}. Nevertheless, in this paper we extend
the WKB asymptotic approach to propose an approximate model that can be solved analytically to give the final solution.
The dimensionless parameter that represents the dielectric mismatch is clearly visualized in the form of the asymptotic solution
and shall provide more intuitive insights for physical understandings.

The WKB approximation for the GDH equation was first introduced by Buff and Stillinger \cite{Buff:JCP:63} for one-plate
problem. The WKB method solves the Green's function problem for a salt-free system by the image-charge theory,
and then approximates the salty system by using screened Coulomb potential to replace each Coulomb potential
where the inverse screening parameter is given by the average between $x$ and $x'$, i.e.,
$$\kappa'(x,x')=\frac{1}{x-x'} \int_{x'}^x \kappa(s) ds.$$
This method was later extended to solve two-plate problem \cite{XML:PRE:2014}. For the system we are studying,
the three-layer structure in each electrode results in a difficulty for an image theory.
To derive the WKB approximation, one starts from the following original Debye-H\"uckel equation of constant coefficient,
\begin{equation}
	-\nabla^2  \widetilde{G} + {\kappa}^2  \widetilde{G} = 4\pi q \delta(\bm{r}-\bm{r}'),
\end{equation}
for $|x|<1-a$, and $-\nabla^2  \widetilde{G} =0 $ otherwise, where $\kappa$ is the inverse Debye length which is
a constant approximating $\sqrt{I}/\epsilon$.

Though our treatment is believed to be useful for a more general dielectric profile, we focus on the effect of $\eta_b$
and consider a uniform dielectric permittivity in the electrolyte, i.e., one sets $\eta_s=1$ to avoid tedious derivation.
Thus, the Born energy in Eq. \eqref{eq:co} is constant and can be normalized, and one can write the general solution of $\widetilde{G}$ with far-field boundary
conditions in terms of the following integrals,
\begin{equation}
	\frac{\widetilde{G}}{q}  =
	\begin{cases}
	\begin{array}{l}
		 \int_0^\infty {C_l(\omega,x')e^{\omega x}} J_0(\rho\omega) d\omega, \quad x<-1, \\
		 \int_0^\infty {\left[ D_l(\omega,x')e^{\omega x}+E_l(\omega,x')e^{-\omega x}\right ]} J_0(\rho\omega) d\omega, \quad -1< x<-1+a, \\		
		 \int_0^\infty \frac{e^{-\tau(\omega) |x-x'|}+A(\omega,x')e^{\tau(\omega) x}+B(\omega,x')e^{-\tau(\omega) x}}{\tau(\omega)} J_0(\rho\omega)\omega d\omega, \quad  |x| <1-a, \\	
		 \int_0^\infty {\left[D_r(\omega,x')e^{\omega x}+E_r(\omega,x')e^{-\omega x}\right]} J_0(\rho\omega) d\omega, \quad 1-a< x<1,\\		
		\int_0^\infty {C_r(\omega,x')e^{-\omega x}} J_0(\rho\omega) d\omega, \quad x> 1,
	\end{array}
	\end{cases}
\end{equation}
where $J_0$ is the Bessel function, $\tau(\omega) =\sqrt{\omega^2+{\kappa}^2}$
and  $\rho = \sqrt{(y-y')^2+(z-z')^2}$. The coefficients $A$, $B$, $C_l$, $C_r$, $D_l$, $D_r$, $E_l$ and $E_r$ are solved
with continuities of electric potentials and displacements on $x=\pm (1-a)$ and $x=\pm 1$. Specifically, for the domain of interest,
i.e., $|x|<1-a$, we have
\begin{align}
	A(\omega,x') & =  \frac{-e^{\tau x'}+f_1(\omega)e^{-\tau (x'+2-2a)}}{e^{2\tau (1-a)}/f_1(\omega) - f_1(\omega)e^{-2\tau (1-a)}},\\
	B(\omega,x') & = \frac{-e^{-\tau x'}+f_1(\omega)e^{\tau (x'-2+2a)}}{e^{2\tau (1-a)}/f_1(\omega) - f_1(\omega)e^{-2\tau (1-a)}},
\end{align}
where the dielectric boundary effect has been embedded into $f_1$ through the dimensionless parameter $\gamma = (1-\eta_b)/(1+\eta_b)$, i.e.,
$$f_1(\omega)=\frac{\omega-\tau {\left(e^{2\omega a}+\gamma\right)}/{\left( e^{2\omega a}-\gamma\right)}}{\omega+\tau {\left(e^{2\omega a}+\gamma\right)}/{\left( e^{2\omega a}-\gamma\right)}}.$$

Consistently, the referenced free-space Green's function is rewritten into the integral form,
\begin{equation}
	\frac{1}{|\bm{r}-\bm{r}'|} = \int_0^\infty e^{-\omega |x-x'|}J_0(\rho\omega) d\omega,
\end{equation}
and then we obtain  the difference in the limit {that is further defined as the rescaled electrostatic correlation energy $u_\textup{el}$, i.e.,}
\begin{align}
	u_\textup{el}= &\lim_{\bm{r}'\rightarrow\bm{r}} \left({\widetilde{G}}/{q}-{1}/{|\bm{r}-\bm{r}'|}\right)  \nonumber\\
	=& \lim_{\rho\rightarrow 0,\ x'\rightarrow x} \left[ \widetilde{G}(x,x',\rho)- \int_0^\infty e^{-\omega |x-x'|}J_0(\rho\omega) d\omega\right] \nonumber\\
	 = &\lim_{x'\rightarrow x}\int_0^\infty \left(\frac{\omega e^{-\tau|x-x'|}}{\tau} -e^{-\omega|x-x'|}\right) d\omega
	 + \int_0^\infty\frac{2f_1e^{-2\tau(1-a)}-e^{2\tau x}-e^{-2\tau x}}{e^{2\tau(1-a)}/f_1-f_1e^{-2\tau(1-a)}}\frac{\omega}{\tau}d\omega \nonumber\\
	=& -\kappa  \left[1-\int_1^\infty \frac{2f_2e^{-2\kappa (1-a)t}-e^{2\kappa  xt}-e^{-2\kappa  xt}}{e^{2\kappa (1-a)t}/f_2-f_2e^{-2\kappa (1-a)t}} dt \right], \label{eq:wkb}
\end{align}
where $f_2(t) = f_1\left(\kappa\sqrt{t^2-1}\right)$.

The WKB approximation for a variable screening is simply to replace the constant $\kappa$ in Eq. \eqref{eq:wkb}
by function $\kappa(x)$. This treatment is asymptotically correct at both low and high screening limits.
Obviously, the WKB approximation has considered the effects of the local ionic strength, the Stern layer
and the dielectric boundary that are dominating unless the variation of ionic strength is quite significant. Thus, the WKB can generally produce well accuracy in solving the GDH, as demonstrated in
Fig. \ref{FDM_INT} in Sec. \ref{num}. It is remarked that Ref. \cite{XML:PRE:2014} studied the WKB approximation
for two-plate problems where the Stern layer effect is not taken into account, and the new WKB is supposed to
provide a better approximation.

\subsection{Numerical method for the modified PNP system}
Though many asymptotic works have been done for the PNP or modified systems \cite{Chen97SIAM,Bazant05SIAM,KBA:PRE:2007,LTZ:2012:JDDE}, the HS interactions in current model cause great challenges in any asymptotic analysis. Instead, we introduce full numerical methods for the nonlinear coupled equation system. As stated above, the excess chemical potentials $\mu^{\textup{co}}_i$ and $\mu^{\textup{hs}}_i$ are functions of $c_i$ only thus can be evaluated by independent numerical scheme at each time step. The main PNP system is then solved by proper schemes and serves to produce updated ion densities $c_i$ and electrostatic potential $\phi$.

For simplicity, we consider the binary electrolyte with symmetric size in this work, i.e, $a_1=a_2=a$  though our numerical schemes can be extended for general asymmetrical electrolytes. On the similar basis, we consider the boundary potentials as $V_+=-V_-=V$.
The main PNP equations are solved numerically on the domain $|x|<1-a$. We introduce uniform half grid points $x_{n+1/2}=(-1+a)+nh$, $n=0,\cdots, N$,
with spatial grid size $h=2(1-a)/N$ such that $x_{N+1/2}=1-a.$ Also, we define the integer grid points $x_n$ as the average of two neighboring
half grid points with two additional ghost points $x_0 = x_1-h$ and $x_{N+1}=x_N+h$.
The uniform time steps are defined by, $t_m = (m-1)k,\ m=1,2,...$, with $k$ the uniform time step size.

For given ionic densities $c_i^{n,m}$ at time $t=t_m$ and position $x_n$, the electric potential is solved approximately by the discretization of the Poisson's equation through a second-order central difference scheme,
\begin{equation}
	-2\epsilon^2\frac{\phi^{n+1,m}-2\phi^{n,m}+\phi^{n-1,m}}{h^2} = \sum_i z_ic_i^{n,m},
\end{equation}
with Robin boundary conditions on $x=\pm (1-a)$ approximated by,
\begin{align}
	\left(\frac{1}{2}+\frac{a}{h}\right)\phi^{0,m} + \left(\frac{1}{2}-\frac{a}{h}\right)\phi^{1,m} &= V_-,\\
	\left(\frac{1}{2}-\frac{a}{h}\right)\phi^{N,m} + \left(\frac{1}{2}+\frac{a}{h}\right)\phi^{N+1,m} &= V_+.
\end{align}
At next time step $t=t_{m+1}$, the ionic density is updated by the discretization of the modified NP equation
\begin{equation}
	\partial_tc_i = \epsilon\partial_x\left(e^{-U_i}\partial_x g_i \right),
\end{equation}
where the generalized Slotboom variables $g_i = c_ie^{U_i}$ is utilized with recalling that $U_i = z_i\phi+\mu^{\textup{co}}_i + \mu^{\textup{hs}}_i$.
Following the strategies in \cite{Liu18SIAM,Ding19JCP}, we discretize the modified NP equation as follows,
\begin{align}
	\frac{c_i^{n,m+1}-c_i^{n,m}}{k} = &\frac{\epsilon}{h}\left(  e^{-U_i^{n+1/2,m+1/2}}\frac{g^{n+1,m+1/2}_i-g^{n,m+1/2}_i}{h} \right.  \nonumber \\	
	& \left.- e^{-U_i^{n-1/2,m+1/2}}\frac{g^{n,m+1/2}_i-g^{n-1,m+1/2}_i}{h}   \right),
\end{align}
with $	g^{n,m+1/2}_i = c^{n,m+1/2}_ie^{-U^{n,m+1/2}_i},$
where proper approximations of averaging are given by
\begin{align}
	c^{n,m+1/2}_i &= \frac{c_i^{n,m+1}+c_i^{n,m}}{2},  \label{eq:av1}\\
	U^{n,m+1/2}_i &= \frac{3U_i^{n,m}-U_i^{n,m-1}}{2},  \label{eq:av2}\\
	U^{n+1/2,m}_i &= \frac{U_i^{n+1,m}+U_i^{n,m}}{2}.  \label{eq:av3}
\end{align}
The no-flux boundary conditions lead to
\begin{align}
	\frac{g^{1,m+1/2}_i-g^{0,m+1/2}_i}{h} =\frac{g^{N+1,m+1/2}_i-g^{N,m+1/2}_i}{h} = 0.
\end{align}
Note that the averages in Eqs. \eqref{eq:av2} and \eqref{eq:av3} have made all nonlinear terms including $\mu^{\textup{co}}_i$
and $\mu^{\textup{hs}}_i$ evaluated explicitly, and the implicit scheme from Eq. \eqref{eq:av1} leads to a final linear system
that {can be solved directly}. This scheme has second order of accuracy and is stable under the condition $k={O}(h)$.
Simultaneously, it is not difficult to prove that the mass conservation is satisfied automatically with time evolution.

\section{Numerical results and discussion} \label{num}

In this section, we {first validate the numerical schemes and then demonstrate the correctness of our theoretical model 
by comparing to particle-based simulation results.}
To validate the accuracy in computing HS interactions, uniform ionic densities $c_i(x)\equiv 1$ are used for
the evaluation of excess HS chemical potentials with different $N$. Remaining parameters take $(\epsilon, q, a) = (0.2, 0.3, 0.15)$.
The results are displayed in Fig. \ref{fig:hs}. We observe the convergence with the increase of $N$ in panel (a),
and that the second order of accuracy is shown in panel (b) as expected for the linear interpolation utilized in
the evaluation of integrals. Note that in panel (b) the errors are differences between numerical results and
analytical solutions of $\mu^{\textup{hs}}(x=0)$.

 \begin{figure}[h!]
\centering
 \includegraphics[width=0.9\linewidth]{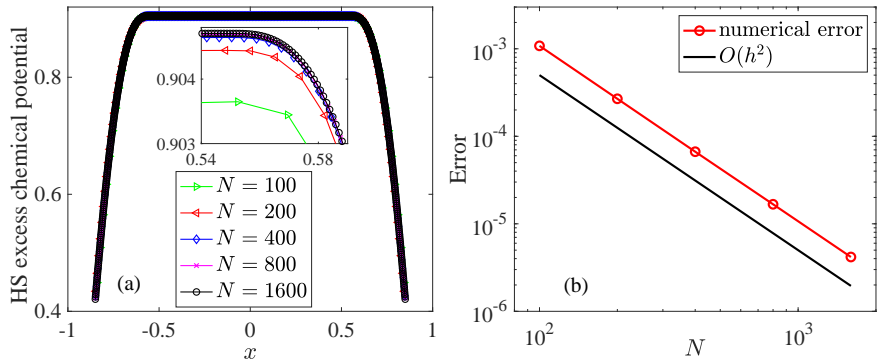}
 \caption{Numerical convergence of computing the HS interactions: {(a) chemical potential for different $N$};
  (b) Errors vs. $N$. Uniform ion density profiles $c_i(x)\equiv 1$ are assumed. }
\label{fig:hs}
 \end{figure}

To show the accuracy of the proposed WKB approximation,
the comparison of solutions of the modified PNP equations is presented in Fig. \ref{FDM_INT} where both the direct finite difference method (FDM) \cite{XuMaggs:JCP:14,XML:PRE:2014}
and the WKB are used in solving the GDH. The dimensionless dielectric constant of the plates is set to be
$\eta_b=0.05$ {where the repulsive effect on ion densities is expected}.
Other parameters chosen for this example are $(\epsilon, q, a,V)=(0.1,0.15,0.075,1)$, $z_1=-z_2=1$, $N=800$ and $k/h=1/2$.
The density profiles are generally in good agreements except a slight deviation in a longer time simulation ($t=5$ in panel (b)) {and both methods can capture the repulsive effect near the surface.}

 \begin{figure}[h!]
 \centering
 \includegraphics[width=0.95\linewidth]{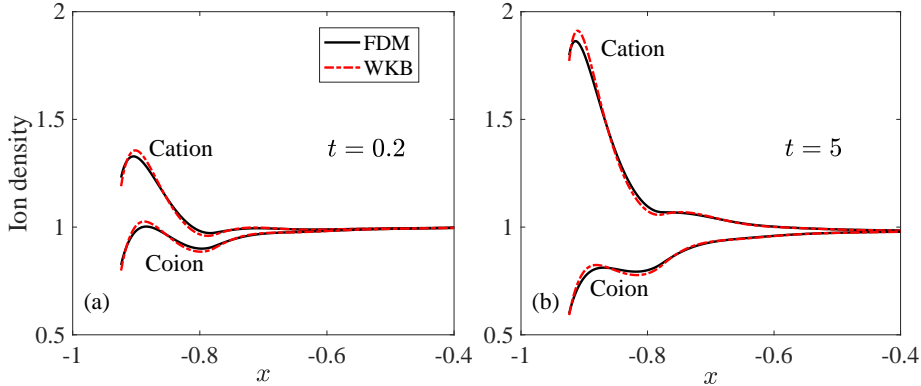}
 \caption{Ionic density profiles from the solution of the modified PNP equations with a comparison between FDM and WKB for solving the GDH equation.}
 \label{FDM_INT}
 \end{figure}

{We next calculate the ion densities from the modified PNP model in comparison to the particle-based MC or MD simulations. We also calculate results from three different models for comparison, including the classical mean-field PNP, the modified PNP with only short-range  HS correlation, the modified PNP with long-range Coulomb correlation, and the modified PNP proposed in current work with both long- and short-range correlations. For short, they are called MF, SC, LC and LS, respectively. As only equilibrium state is needed here, we take the Boltzmann distribution for ion densities expressed as $c_i=\exp(z_i\phi+\mu_i^{\textup{co}} + \mu_i^{\textup{hs}})$ and the resulting modified PB equation system is discretized by the FDM and solved iteratively \cite{XuMaggs:JCP:14,Ma14JCP}. We start from the cases without dielectric mismatch on surfaces, i.e., $\gamma=0$, and the results are presented in Fig. \ref{MC} (a) where the MC simulation results are adopted from recent work \cite{valisko2018systematic}. 
{For uncharged surfaces, the LS, LC and MC results predict the slight repulsion of ions near the surface due to the ionic depletion effect beyond the MF theory. Additionally,  the agreement of MC and theoretical results (LS, LC) for the case with weakly charged surfaces ($0.02\ C/m^2$) is observed. Due to the weak HS correlation effect here, there is a very slight difference between the LS and LC models and the weak attraction near the uncharged surfaces is observed from SC model, which are similar to those in panels (b) and (c).
In Fig. \ref{MC} (b), the cases of one single dielectric surface with high and low dielectric constants are studied. Our theoretical predictions from the LS and LC match the MC results \cite{boda2004computing,henderson2005monte} well except a slight deviation approaching the dielectric-repulsive surface. In Fig. \ref{MC} (c), we calculate the system with two uncharged but dielectric surfaces. For both dielectric-repulsive and attractive surfaces, the LS and LC results agree well with MD simulations from the recent work\cite{liang2020harmonic}.} 
We further study the case with significant HS correlation effects in panel (d) with uncharged surfaces, where the dimensionless parameters are $(\epsilon,q,a,\gamma)=(0.2,0.3,0.15,0.95)$ and the MD simulation is performed with the method in Ref. \cite{liang2020harmonic}.
Large deviations from results of the four different models are depicted. The SC model predicts strong adsorption of ions near the surface 
due to the HS correlation, namely, particle repulsion in the interior region is stronger than those near the surface. In contrast, the LC model only predicts a rapid decline of ion density caused by the dielectric repulsion. As both long and short effects are competitive, the LS model produces the peak of density near the surface, which agrees well with the MD simulation. Note that there is an increased deviation approaching to the surface.
This is due to the Lennard-Jones treatment in MD,  instead of the hard spheres in the LS model.
The detailed information for all corresponding dimensional parameters can be found in figure caption.
}
   \begin{figure}[t!]
 \centering
 \includegraphics[width=4.8in]{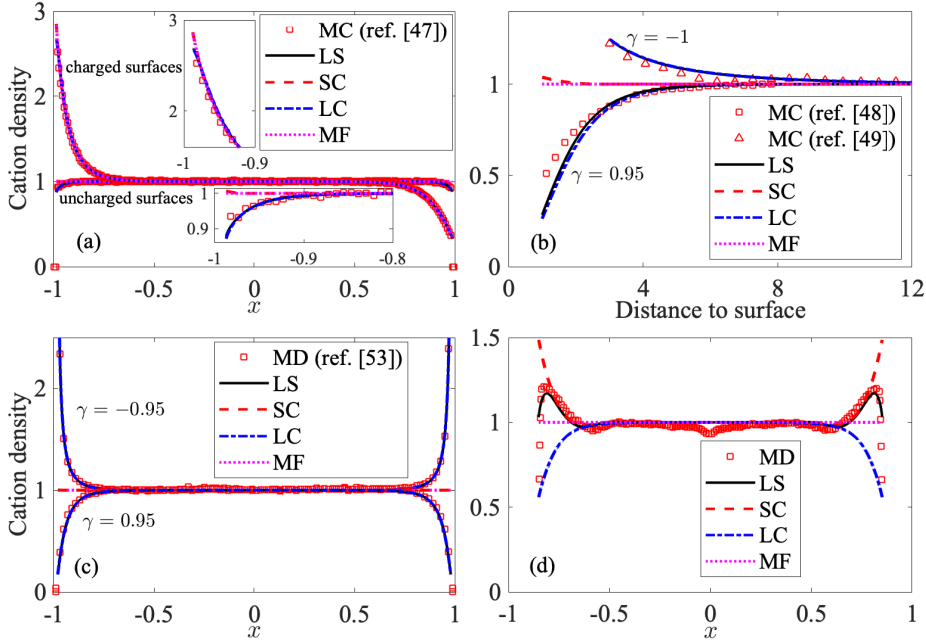}
 \caption{Cation density profiles for $1:1$ electrolyte. \textup{(a)}  Two parallel surfaces without dielectric mismatch. MC results are from the free database of \cite{valisko2018systematic}. The dimensional parameters are $c_0=100$ mM, $a_\pm=0.15$ nm, $L=11.825$ nm.  For the charged case, the surface charge densities are $\mp0.02\ C/m^2$ on the left and right surfaces.  The insets show profiles more clearly near the surfaces. \textup{(b)} One single uncharged dielectric surface with the distance to the surface in the unit of the ion radius $a_\pm=0.15$ nm. MC results of the $\gamma=-1$ case are from \cite{henderson2005monte} (solid squares in their figure 3) with $c_0=50$ mM, and those of the $\gamma=0.95$ case are from \cite{boda2004computing} (in their figure 4 (c)) with $c_0=500$ mM. \textup{(c)} Two dielectric surfaces. MD results are from \cite{liang2020harmonic} (in their figure 5 left) where $c_0=90$ mM, $L=5$ nm and $a_\pm=0.1$ nm. \textup{(d)} Two dielectric surfaces with 
 parameters $\gamma=0.95$, $c_0=424$ mM, $L=2.33$ nm and $a_\pm=0.35$ nm and simulation results are obtained by MD  \cite{liang2020harmonic}.
  The Bjerrum length $\ell_\textup{B}$ is $0.714$ nm in \textup{(ab)} and $0.7$ nm in \textup{(cd)}.}
 \label{MC}
 \end{figure}

   \begin{figure}[h!]
 \centering
 \includegraphics[width=0.95\linewidth]{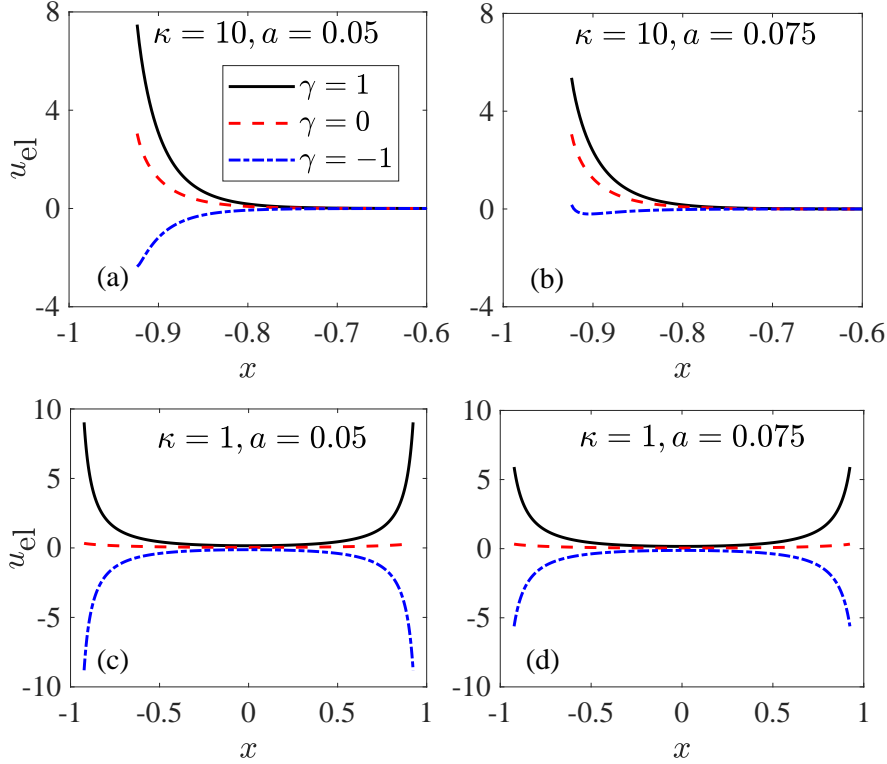}
 \caption{The rescaled electrostatic correlation energy (defined in Eq. \eqref{eq:wkb})
 with different dielectric ratios and a homogeneous ionic strength.}
 \label{fig2}
 \end{figure}

In the rest of the paper, three cases of parameter $\gamma=1, 0, -1$ are considered. Our WKB results show that
the dielectric effect is enhanced with the decrease of $\kappa a$ which corresponds to a more dilute electrolyte and/or
smaller ionic size (thinner Stern layer thickness). As a simple case to validate the model, we assume the artificial
environment with homogeneous ionic strength, $\kappa(x)\equiv\kappa$, and test {the rescaled electrostatic correlation energy $u_\textup{el}$ \eqref{eq:wkb}} 
with $\kappa=10, 1$ and $a=0.05, 0.075$, where effects of dielectric boundaries are  well visualized in Fig. \ref{fig2}.

  \begin{figure}[h!]
\centering
 \includegraphics[width=0.95\linewidth]{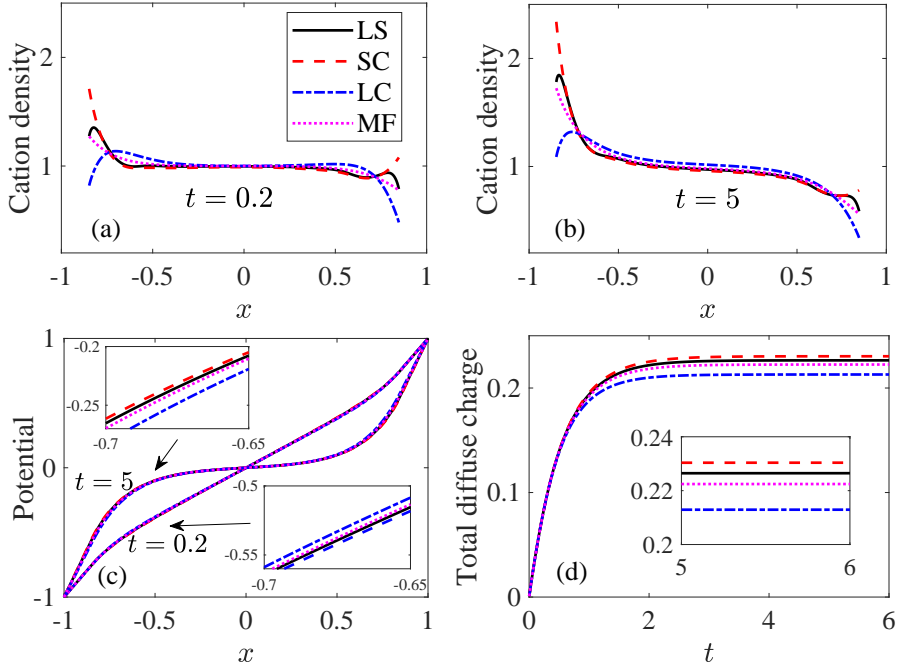}
 \caption{Cation density, potential and total diffuse charge profiles for four different models.}
\label{fig:effect_2}
 \end{figure}

We next simulate the modified PNP system by studying the 1:1 binary electrolyte with fixing $V=1$ and varying the
dimensionless parameters $(\epsilon, q, a, \gamma)$.   Fig. \ref{fig:effect_2} shows the results with
parameters $(\epsilon, q, a, \gamma) = (0.2, 0.3, 0.15, 1)$.
The cation densities are shown in panels (a) and (b) at time $t=0.2$ and $t=5$, where the overlapping of
electric double layers on both interfaces has been observed due to the relatively large value of $\epsilon$.
Compared with MF, the SC model enhances the density profile obviously while the LC model causes the repulsive effect
because of the high dielectric ratio $\gamma = 1$ which means the low dielectric permittivity of the boundary materials.
The density profile from the LS model is close to that from the SC model away from the surface, but the deviation is
increased as approaching the Stern layer due to the repulsive effect. Both the depletion and the layering effects
are observed in the LS model in agreement with the physical intuition of the double-layer structure, demonstrating
the importance of including both the long-range and short-range correlations in the modified PNP theory.

Fig. \ref{fig:effect_2} (c, d) displays potential and total diffuse charge profiles from the four different models.
The total diffuse charges present a clear discrepancy in the four models when the system tends to the steady state at long time.
This is due to the opposite effects between the Coulomb correlation and HS correlation, where
{the former decreases the diffuse charge and the latter improves it. Therefore, the model with only the long-range
correlation underestimates the total diffuse charge, and that with only the HS correlation overestimates
it.}
 \begin{figure}[h!]
\centering
 \includegraphics[width=0.95\linewidth]{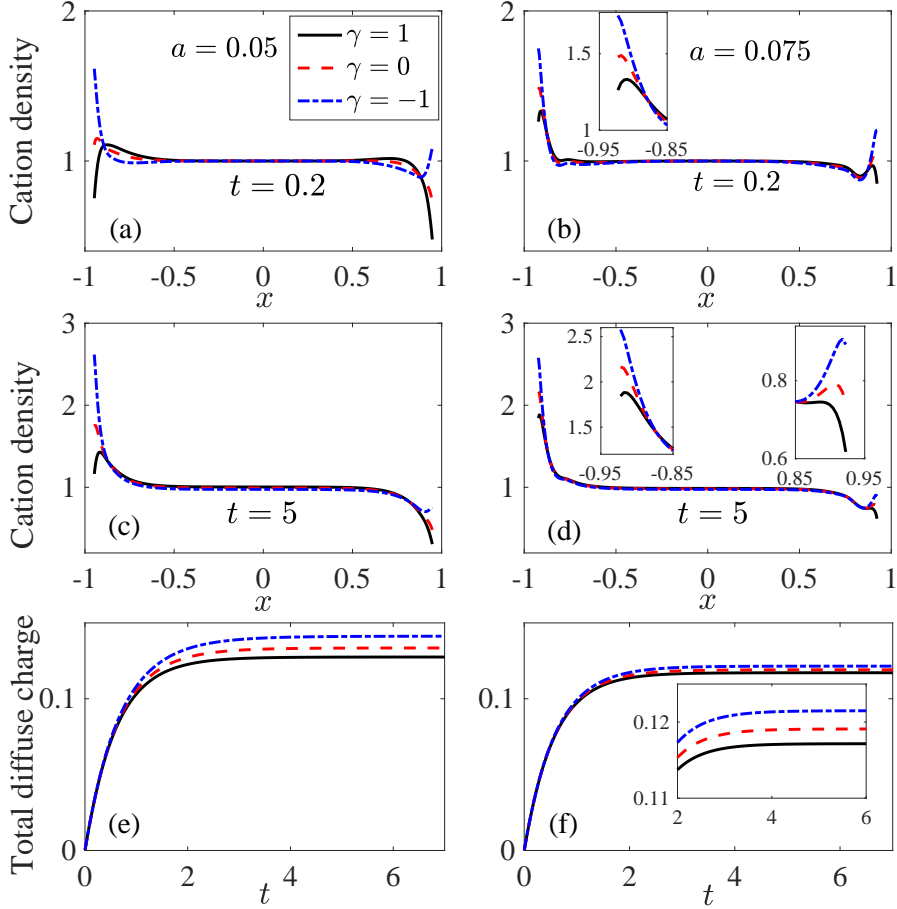}
 \caption{Cation density, potential and diffuse charge profiles with different dielectric ratios for ionic size \textup{(ace)} $a=0.05$
 and \textup{(bdf)} $a=0.075$.}
\label{fig:effect_3}
 \end{figure}

To have a closer look at mutual effects of dielectric boundaries and HS interactions, we study the solutions of the LS
model by varying $\gamma$ and $a$ with fixed $(\epsilon, q, V)$. In Fig. \ref{fig:effect_3}, we present the cation density
and diffuse charge for $a=0.05$ and $0.075$ and different $\gamma$ values. The deviations for different $\gamma$ are larger
for a smaller size $a$ (a, c) because of  the reduced effect of HS interactions in comparison to the dielectric effect
and the thinner Stern layer thickness enhancing the dielectric boundary effect. When $a$ increases (b, d), the density oscillating
behavior starts to occur and the deviations in density profiles with different $\gamma$ are reduced. From panels (e, f),
the attractive effect in the case of high electrode permittivity ($\gamma=-1$) enhances the total diffuse charge.
Similarly, the deviations in diffuse charges decrease for larger $a$ due to the reduced dielectric effect.

\section{Conclusion}\label{con}
By taking into account both long-range Coulomb and short-range HS correlations through an energetic approach,
we have made proper improvements to the classical PNP model and developed a self-consistent modified PNP model including
both correlations.  The Coulomb correlation energy is obtained by solving the GDH equation, and the HS interaction
is captured accurately by the MFMT. We have demonstrated the significant improvements of our modified PNP by comparing
it with mean-field or other partially modified models through the two-plate problem with a piecewise-constant dielectric
profile. An accurate and efficient hybrid numerical method, which incorporates the WKB approximation for GDH
into a stable and mass-conservative scheme for the main PNP, has been developed to solve the entire two-plate modified
PNP model.
By calculating the results for systems with different values of $\gamma$ and $a$, we have explored the competitive effects from
the dielectric ratio and ionic size, which is essentially the main motivation of the current work.

The present modified PNP model serves as a powerful approach beyond the mean-field treatment considering that both long- and short-range
correlations play the same important role in many practical systems at the micro/nano scale. {As one of the possible applications, the phenomena of the like-charge attraction or charge inversion is worth to be explored by the current model. The competition between the surface charge and the dielectric repulsive effect has been considered to lead to like-charge attraction or charge inversion \cite{wang2013effects}.  Recently, a general theory is developed to consider the effects of strong ion-ion correlations and short-range hydration interactions on surface forces \cite{misra2019theory}. A systematic study with surface charge, long-range electrostatic correlation (including dielectric boundary effect) and short-range HS interaction on these many-body phenomena will be of much interest and will be the goal of future work.}

\section*{Acknowledgement}
The authors thank Jiuyang Liang for the help on preparing the MD simulation results.
M. M. acknowledges the financial support from the Natural Science Foundation of China (NSFC, Grant No. 11701428), ``Chen Guang'' project
by Shanghai Municipal Education Commission and Shanghai Education Development Foundation, and the Fundamental Research Funds for the Central Universities.
Z. X. and L. Z. acknowledge the financial support from the NSFC (Grant No. 12071288), the Shanghai Science
and Technology Commission (grant No. 20JC1414100) and the support from the HPC center of Shanghai Jiao Tong University.
Part of the work was completed during M.M.'s visit to the University of Melbourne.

\medskip


\end{document}